\documentstyle[preprint,aps,prb,psfig]{revtex}

\draft

\begin{document}
\title{Dynamic critical properties of the vortex--glass transition
derived from angular-dependent properties of
La$_{2-x}$Sr$_{x}$CuO$_{4}$ films}

\author{T.\ Schneider$^{a}$, G.\ I.\ Meijer$^{b}$, J.\
Perret$^{c,b}$, J.-P.\ Locquet$^{b}$, and P.\ Martinoli$^{c}$}

\address{$^{a}$Physik-Institut, Universit\"at Z\"urich,
Winterthurerstr.\ 190, 8057 Zurich, Switzerland
\\
$^{b}$IBM Research, Zurich Research Laboratory, 8803 R\"uschlikon, Switzerland
\\
$^{c}$Institut de Physique, Universit\'e de Neuch\^atel, 2000 Neuch\^atel, Switzerland}
\maketitle

\begin{abstract}
We present resistivity data on a high-quality
La$_{2-x}$Sr$_{x}$CuO$_{4-\delta }$ film measured in a magnetic
field of 1 T applied at an angle $\delta$ to the $c$ axis. Using these data,
the influence of the orientation of the
magnetic field and the effective mass anisotropy on the
vortex--glass transition can be studied. The variation of $\delta$ for a fixed
magnitude of the magnetic field allows us to investigate the
critical properties of interest, including the 2D-to-3D crossover
and the 3D vortex glass-to-fluid transition, as the temperature is
decreased. The data are well described by the scaling theory for
the d.c.\ resistivity of an anisotropic superconductor in a
magnetic field applied at an angle $\delta$ to the $c$ axis. This
scaling includes the critical properties close to and at the
vortex--glass transition. The main results include\ (i) evidence
of a Kosterlitz--Thouless transition in zero field, and (ii) a
2D-to-3D crossover at $H=1$ T as the temperature is decreased
below the zero-field transition temperature, leading to the
vortex fluid-to-vortex glass transition in D = 3 characterized by
the dynamic critical exponent $z\approx 5.7$.
\end{abstract}

\pacs{05.70.Jk, 74.25.Fy, 74.40.+k}

\clearpage

\section{Introduction}

\label{sec:intro}

The critical dynamics of the vortex system in cuprate
superconductors is strongly affected by the combined effect of
pinning, thermal fluctuations, anisotropy, and
dimensionality.\cite{Blatter94} Thermal fluctuations are
responsible for the existence of a first-order vortex lattice
melting transition in clean systems. In the presence of disorder,
however, the long-range order of the vortex lattice is destroyed
and the vortex solid becomes a glass. The vortex fluid-to-glass
transition appears to be a second-order transition, signaled by
the vanishing of the zero-frequency resistance in the vortex-glass
phase. A schematic sketch of the phase diagram is shown in Fig.\
\ref{fig1}. By lowering the magnetic field at temperatures below $T_{c}$
($H=0$) the material undergoes the vortex liquid-to-vortex glass
or Bose glass transition at $H=H_{g}$ (path 3). Similarly, at
constant field, the transition occurs at $T_{g}<T_{c}$ (path 2)
for lower temperatures.

In the Anderson--Kim flux-creep model,\cite{Anderson64} and extensions
thereof\cite{Tinkham88} that include pinning effects, the
collective effects of vortices are neglected. In these models,
the linear resistance is predicted to drop rapidly upon cooling
(path 2) but always remains nonzero. On the other hand, there is
considerable evidence of a nearly continuous vortex melting
transition. In the presence of pinning, however, Larkin and
Ovchinnikov \cite{Larkin79} have shown that the long-range order
of the vortex lattice is destroyed in all spatial dimensions
smaller than four. Hence in two and three dimensions pinning
should dominate. Following the work of Ebner and
Stroud,\cite{Ebner85} Fisher\cite{Fisher89} and Fisher, Fisher and Huse\cite{Fisher91a}
have postulated a continuous vortex--glass
transition with infinite conductivity in the glass phase.
Experimental evidence of this transition has been obtained from
the critical behavior of transport properties for magnetic fields
applied parallel to the $c$
axis.\cite{Koch89,Gammel91,Kotzler94,Woltgens95,Mun96,Rydh99,Sefrioui99}

In this study, we consider the full range of the angle $\delta$
between the applied magnetic field and the $c$ axis. We present
resistivity data on a high-quality La$_{2-x}$Sr$_{x}$CuO$_{4}$
film with $x\approx 0.1$ in a magnetic field of 1 T applied at an
angle $\delta$ to the $c$ axis. This allows us to study the
influence of the orientation of the magnetic field and the
effective mass anisotropy on the glass transition. By varying
$\delta$ for a fixed magnitude of the magnetic field, we have
been able to investigate the critical properties of interest,
including the 2D-to-3D crossover and the 3D vortex glass-to-fluid
transition.
Noting that in this film disorder is much stronger than in high-quality
single crystals, we do not consider the Bragg--glass transition.\cite{Banerjee}

The paper is organized as follows. In Sec.\ \ref{sec:scaling} we
first summarize the scaling theory for the d.c.\ resistivity of an
isotropic superconductor. This formalism is then extended to
account for anisotropy and an angle $\delta$ between magnetic
field and the $c$ axis. The scaling includes the critical
properties close to and at the vortex--glass transition. Section
\ref{sec:bulk} contains the experimental details, results and
their analysis. The main results include (i) evidence of a
Kosterlitz--Thouless transition in zero field; (ii) a 2D-to-3D
crossover at $H=1$ T as the temperature is lowered below the
zero-field transition temperature, leading to the vortex
fluid-to-vortex glass transition in D = 3 characterized by the
dynamic critical exponent $z_{g}\approx 5.7$.

\section{Scaling Theory}

\label{sec:scaling}

\begin{figure}
\begin{center}
\leavevmode \psfig{figure=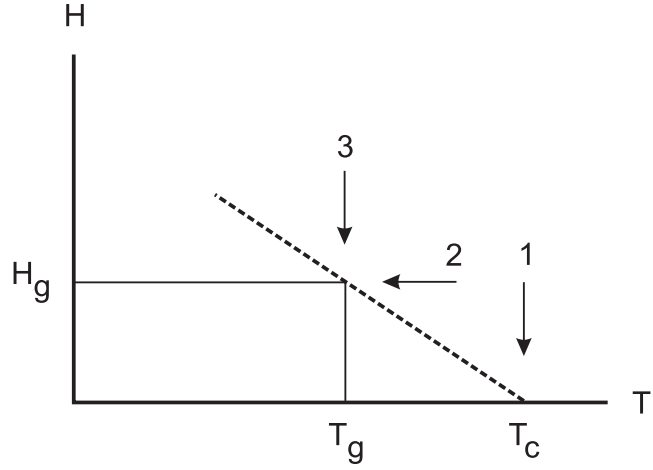,width=85mm}
\end{center}
\caption{
Schematic sketch of the $(H,T)$ phase diagram for the
vortex--glass transition.
}
\label{fig1}
\end{figure}

A continuous phase transition from the superconducting to the
normal phase allows a general formulation of the scaling at and
near such a transition.\cite{Schneider98} In an isotropic system
the fluctuation contribution to the conductivity adopts the
scaling form
\begin{equation}
\sigma ^{F}\propto \xi ^{2+z_{g}-{\rm D}}{\cal F}({\cal Z}),\ \
{\cal Z}= \frac{H\xi ^{2}}{\Phi _{0}},
\label{eq1}
\end{equation}
where $\xi$ is the correlation length, $H$ the magnetic field, $z$
the dynamic critical exponent, and ${\cal F}({\cal Z})$ a
universal scaling function. Its form will be discussed later. Close to the vortex--glass transition
the correlation length diverges with an exponent $\nu_{g}$
as\cite{Fisher89}
\begin{equation}
\xi =\xi _{T,0}\left( \frac{T-T_{g}}{T_{g}}\right) ^{-\nu
_{g}}\text{ and }\ \xi =\xi _{H,0}\left(
\frac{H-H_{g}}{H_{g}}\right) ^{-\nu _{g}}  \label{eq2}
\end{equation}
along paths 2 and 3 (see Fig.\ \ref{fig1}), respectively. Hence it
follows that the conductivity will diverge along the
phase-transition line
\begin{equation}
{\cal Z}={\cal Z}_{g}.
 \label{eq3}
\end{equation}
In this context it is important to recognize that ${\cal Z}_{g}$
is a number that locates the transition. Close to the zero-field
transition of a bulk sample at $T_{c}$ ($H=0$), the isotropic
3D-$XY$ fluctuations are expected to dominate. In this case the
phase-transition line in Fig.\ \ref{fig1} follows from
\begin{equation}
{\cal Z}_{g}=\frac{H_{g}\xi ^{2}(T)}{\Phi _{0}},\ \ \xi (T)=\xi
_{0}\left( \frac{T_{c}(H=0)-T}{T_{c}(H=0)}\right) ^{-\nu },\;\nu
\approx 2/3 \, ,
\label{eq4}
\end{equation}
so that
\begin{equation}
H_{g}=\frac{{\cal Z}_{g}\Phi _{0}}{\xi _{0}^{2}}\left(
\frac{T_{c}(H=0)-T}{ T_{c}(H=0)}\right) ^{2\nu }.  \label{eq5}
\end{equation}
At ${\cal Z}_{g}$ the scaling function exhibits singular behavior,
signaling the occurrence of the transition. Its critical behavior
is governed by the glass correlation length [Eq.\ (\ref{eq2})]
rewritten as
\begin{equation}
\xi \propto \left( \frac{T-T_{g}}{T_{g}}\right) ^{-\nu
_{g}}\propto \left( \frac{H-H_{g}}{H_{g}}\right) ^{-\nu
_{g}}\propto \left( {\cal Z}-{\cal Z} _{g}\right) ^{-1/2}.
\label{eq6}
\end{equation}
Combining Eqs.\ (\ref{eq1}), (\ref{eq3}) and (\ref{eq6}), we find
that the conductivity will diverge along path 2 (see Fig.\ 1) as
\begin{equation}
\sigma ^{F}\propto t^{-\nu _{g}(2+z-D)}{\cal F}({\cal Z}_{g}),\ \
t=\frac{ T-T_{g}}{T_{g}}  \label{eq7}
\end{equation}
 and along path 3 (Fig.\ 1) as
\begin{equation}
\sigma ^{F}\propto \left( \frac{H-H_{g}}{H_{g}}\right) ^{-\nu
_{g}(2+z_{g}-D)}{\cal F}({\cal Z}_{g}) . \label{eq8}
\end{equation}
However, along path 1, where $T=T_{c}(H=0)$, the conductivity is
finite for any $H\neq 0$. Rewriting the conductivity [Eq.\
(\ref{eq1})] in the form
\begin{equation}
\sigma ^{F}\propto \xi ^{2+z-D}{\cal F}({\cal Z})=\left(
\frac{{\cal Z}\Phi _{0}}{H}\right) ^{(2+z-D)/2}{\cal F}({\cal Z})
, \label{eq9}
\end{equation}
we see that it scales as
\begin{equation}
\sigma ^{F}\propto \left( \frac{\Phi _{0}}{H}\right)
^{(2-D+z)/2},\ \ {\cal F }({\cal Z})\propto {\cal Z}^{-(2+z-D)/2}.
\label{eq10}
\end{equation}
The dynamic critical exponent $z$ is expected to differ from that
entering the normal-to-vortex glass transition ($z_{g}$). Indeed,
as $H$ is lowered one approaches in this case the zero-field
transition point.

At the normal-to-vortex glass transition, the conductivity is
supposed to be infinite. This requires the scaling form
\begin{equation}
\sigma ^{F}\propto ({\cal Z}-{\cal Z}_{g})^{-(2+z_{g}-D)/2}\ .
\label{eq11} \end{equation}
Indeed, according to Eqs.\ (\ref{eq1}) and (\ref{eq6}) we have
\begin{equation}
\sigma ^{F}\propto \xi ^{2+zg-D}{\cal F}({\cal Z}),\ \ \ \ \xi
\propto ( {\cal Z}-{\cal Z}_{g})^{-1/2} , \label{eq12}
\end{equation}
so that close to criticality (${\cal Z}={\cal Z}_{g})$,
\begin{equation}
\sigma ^{F}\propto ({\cal Z}-{\cal Z}_{g})^{-(2+z_{g}-D)/2}{\cal
F}({\cal Z}_{g}).  \label{eq13}
\end{equation}
For anisotropic superconductors and in particular for arbitrary
orientation of the applied field with respect to the $c$ axis,
various expressions have to be modified.

\section{Bulk superconductor (D = 3) with uniaxial anisotropy}
\label{sec:bulk}

For a bulk superconductor with uniaxial anisotropy the scaling
variable adopts the form\cite{Sefrioui99}
\begin{eqnarray}
{\cal Z} &=&{\cal Z}(H,T)\ {\cal Z}(\gamma ,\delta ),\ \ \ \ \ \
\gamma = \sqrt{\frac{M_{\bot }}{M_{\Vert }}}
\nonumber
\\
{\cal Z}(H,T)
&=&\frac{\xi _{\Vert }^{2}H}{\Phi _{0}},\
\nonumber
\\
{\cal Z}
(\gamma ,\delta)&=&\left( \cos ^{2}(\delta )+\frac{1}{\gamma ^{2}}\sin ^{2}(\delta
)\right) ^{1/2}\, ,
\label{eq14}
\end{eqnarray}
where $\delta$ is the angle between the $c$ axis and the magnetic
field, and $\xi _{\Vert}$ is the correlation length in the $ab$
plane (perpendicular to the $c$ axis). Here $\gamma$ measures the anisotropy of the effective mass
parallel ($ab$) and perpendicular ($a$) to the layers. Thus the in-plane
conductivity [Eq.\ (\ref{eq1})] adopts in the bulk the scaling
form
\begin{equation}
\sigma ^{F}\propto \xi _{\Vert }^{z_{g}-1}{\cal F}({\cal Z})=\xi _{\Vert
}^{z_{g}-1}{\cal F}\left( {\cal Z}(H,T)\ {\cal Z}(\gamma ,\delta )\right) .
\label{eq15}
\end{equation}
Hence, close to criticality, the $\delta $-dependent conductivity,
$\sigma ^{F}(\delta )$, measured at constant field $H$ and
temperature $T$ plotted versus ${\cal Z}(\gamma ,\delta )$ should
fall on a single curve. In anisotropic bulk systems this offers
the opportunity to probe the scaling function for fixed
temperature and magnetic field in the domain
\begin{equation}
\frac{{\cal Z}(H,T)}{\gamma }\leq {\cal Z}\leq {\cal Z}(H,T).
\label{eq16}
\end{equation}
Supposing that 3D-$XY$ fluctuations dominate close to
$T=T_{c}(H=0)$, the vortex--glass transition line defined by
${\cal Z=Z}_{g}$ [Eq.\ (\ref{eq3})] is then given by the
angular-dependent expression
\begin{equation}
H_{g}(\delta )=\frac{{\cal Z}_{g}\Phi _{0}}{\xi _{\Vert
,0}^{2}}\left( \frac{ T_{c}(H=0)-T}{T_{c}(H=0)}\right) ^{2\nu
}\left( \cos ^{2}(\delta )+\frac{1}{ \gamma ^{2}}\sin ^{2}(\delta
)\right) ^{-1/2},\ \nu \approx 2/3. \label{eq17}
\end{equation}
We note that for $\delta =0$, $2\nu \approx 4/3$ turned out to be
consistent with the data of Gammel {\it et al.}\ for
YBa$_{2}$Cu$_{3}$O$_{7-\delta }$ single crystals.\cite{Gammel91}
At the normal-to-vortex glass transition, the conductivity is
infinite.\cite{Fisher91a} According to Eq.\ (\ref{eq10}) this
implies
\begin{equation}
\sigma ^{F}\propto ({\cal Z}-{\cal Z}_{g})^{-(z_{g}-1)/2},  \label{eq18}
\end{equation}
where
\begin{eqnarray}
{\cal Z}-{\cal Z}_{g} &=&\left( {\cal Z}(H,T){\cal Z}(\gamma
,\delta )-{\cal Z}_{g}\right)
\nonumber \\
&=&{\cal Z}(H,T)\left(
{\cal Z}(\gamma ,\delta )-\frac{{\cal Z}_{g}}{{\cal Z}
(H,T)}\right)  \label{eq19}
\end{eqnarray}
and hence
\begin{equation}
\sigma ^{F}\propto {\cal Z}(H,T)^{-(z_{g}-1)/2}({\cal Z}(\gamma
,\delta )- \widetilde{{\cal Z}}_{g})^{-(z_{g}-1)/2},\
\widetilde{{\cal Z}}_{g}=\frac{{\cal Z}_{g}}{{\cal Z}(H,T)}.
\label{eq20}
\end{equation}
Thus the plot $\sigma ^{F}(\delta )$ versus ${\cal Z}(\gamma
,\delta )$ allows us to estimate the dynamic critical exponent
$z_{g}$ directly. The prefactor depends on the magnetic field and
temperature.

Along path 1 (see Fig.\ \ref{fig1}) the conductivity adopts a
rather different scaling form. In analogy to Eq.\ (\ref{eq10}),
\begin{equation}
\sigma ^{F}\propto \left( \frac{\Phi _{0}}{H{\cal Z}(\gamma
,\delta )} \right) ^{(z-1)/2},\ {\cal F}({\cal Z})\propto {\cal
Z}^{-(z-1)/2}. \label{eq21}
\end{equation}
It is important to recall that this dynamic critical exponent $z$
is expected to differ from that entering the normal-to-vortex
glass transition $(z_{g})$. Indeed, as $H$ is lowered one
approaches the zero-field transition point.

\section{Thin films}
\label{sec:films}

In thin films of thickness $d$, 2D fluctuations dominate as long as
\begin{equation}
\xi _{c}\equiv \xi _{\bot }\geq d.
\label{eq22}
\end{equation}
Here $\xi _{c}$ denotes the correlation length along the $c$ axis,
which is perpendicular to the film. Approaching the
Kosterlitz--Thouless transition \cite{Kosterlitz73} from above,
the zero-field conductivity scales according to Eq.\ (\ref{eq1})
as
\begin{equation}
\sigma ^{F}\propto \xi ^{z},  \label{eq23}
\end{equation}
where $\xi$ is given by
\begin{equation}
\xi \propto \exp (b(T/T_{c}-1)^{-1/2}),\ \ T>T_{c}.  \label{eq24}
\end{equation}
Accordingly, the resistivity scales as
\begin{equation}
\rho \propto \xi ^{-z}\propto \exp (-zb(T/T_{c}-1)^{-1/2}) , \label{eq25}
\end{equation}
and hence
\begin{equation}
(\ln \rho )^{-2}\propto T/T_{c}-1.  \label{eq26}
\end{equation}
In D = 2 and a finite applied magnetic field, the scaling variable
${\cal Z}$ entering the expression for the conductivity [Eq.\
(\ref{eq1})] adopts the form\cite{Schneider93}
\begin{eqnarray}
{\cal Z} &=&{\cal Z}(H,T)\ {\cal Z}(H,\delta )
\nonumber
\\
{\cal Z}(H,T) &=& \frac{H^{1/2}\xi _{\Vert }}{\Phi _{0}^{1/2}},\
\nonumber
\\
{\cal
Z(} H,\delta )& =& \left( (|\cos (\delta )|+\frac{Hd^{2}}{12\Phi
_{0}}\sin ^{2}(\delta )\right) ^{1/2}.  \label{eq27}
\end{eqnarray}
Accordingly, in a film of thickness $d$, angular-dependent
conductivity measurements allow us to probe the scaling function
for fixed temperature and magnetic field in the domain
\begin{equation}
{\cal Z}(H,T)\left( \frac{Hd^{2}}{12\Phi _{0}}\right) ^{1/2}\leq {\cal Z\leq
Z}(H,T).  \label{eq28}
\end{equation}
Comparing the angular dependence of the scaling variable for
anisotropic bulk systems [Eq.\ (\ref{eq14})] with that of a
sufficiently thin film [Eq.\ (\ref{eq27})], it is seen that they
differ markedly around $\delta =\pi /2$. Indeed, ${\cal
Z}(H,\delta )$ has a cusp at $\delta =\pi /2$, whereas ${\cal
Z}(\gamma ,\delta )$ is smooth. Moreover, ${\cal Z}(H,\delta )$
depends on the magnetic field. The cusp of ${\cal Z}(H,\delta )$
at $\delta =\pi /2$, modulo $\pi$ leads to a V-shaped structure
around these angles in the resistivity and, together with
the magnetic field dependence,  allows an unambiguous determination of the
effective dimensionality.

According to Eq.\ (\ref{eq1}) the conductivity in the
regime where 2D fluctuations dominate [Eq.\ (\ref{eq22})] then adopts the
scaling form
\begin{equation}
\sigma ^{F}\propto \xi ^{z_{g}}{\cal F}({\cal Z})=\xi ^{z_{g}}\
{\cal F} \left( {\cal Z}(H,T)\ {\cal Z}(H,\delta )\right) .
\label{eq29}
\end{equation}
Hence close to criticality and for fixed $H$ and $T$,
angular-dependent conductivity data $\sigma ^{F}(\delta )$ plotted
versus ${\cal Z}(H,\delta )$ should fall on a single curve.
Experiments,\cite{Dekker92} simulations,\cite{Fisher91b} and
rigorous analytic arguments\cite{Nishimori94} revealed, however,
that there is no vortex glass at finite $T$ in D = 2. Even though
$T_{g}=0$, the correlation length diverges as $T\rightarrow 0$,
\begin{equation}
\xi \propto T^{-\nu _{g}},  \label{eq30}
\end{equation}
which leads to observable consequences at finite temperatures.
Moreover, as the transition is at $T=0$, the relaxation has an
activated form and diverges as $T\rightarrow 0$. Formally this
corresponds to $z_{g}=\infty$. To eliminate $z_{g}$ in Eq.\
(\ref{eq29}) one can introduce the barrier $\Delta E(T)$ that a
vortex has to cross in order to move a distance $\xi$. In this context one
conventionally defines the barrier exponent $\psi$
by\cite{Hyman94}
\begin{equation}
\Delta E(T)\propto \xi ^{\psi }\propto T^{-\nu _{g}\psi } ,  \label{eq31}
\end{equation}
in terms of which
\begin{equation}
\frac{1}{\tau }\propto \exp (-\Delta E/T)\propto \exp (-C/T^{1+\nu
_{g}\psi }).  \label{eq32}
\end{equation}
With the definition $\tau \propto \xi ^{z}$ the afore-mentioned
problem $( z_{g}=\infty )$ is resolved and the scaling form
of the conductivity [Eq.\ (\ref{eq29})] reduces to
\begin{equation}
\sigma ^{F}\propto \exp (C\xi ^{\psi +1/\nu _{g}}){\cal F}({\cal
Z})=\exp (C\xi ^{\psi +1/\nu _{g}})\ {\cal F}\left( {\cal Z}(H,T)\
{\cal Z}(H,\delta )\right) .  \label{eq33}
\end{equation}

Along path 1 (see Fig.\ \ref{fig1}), however, 2D-$XY$ fluctuations
are expected to dominate. Here the conductivity exhibits,
analogous to Eqs.\ (\ref{eq10}) and (\ref{eq21}), the power-law
behavior
\begin{equation}
\sigma ^{F}\propto \left( \frac{\Phi _{0}}{H}\right) ^{z/2}{\cal
Z}(\gamma ,\delta )^{-z},\ \ {\cal F}({\cal Z})\propto {\cal
Z}^{-z}.  \label{eq34}
\end{equation}

\section{Results and Discussion}

\label{sec:results}

We have measured the angular dependence of the in-plane
resistivity on a 120-\AA\/-thick La$_{2-x}$Sr$_{x}$CuO$_{4}$
film. The film was grown by molecular-beam epitaxy on a
(001)-oriented LaSrAlO$_{4}$ substrate. Details of the sample
preparation have been described elsewhere.\cite{Locquet98} The
angle-dependent resistivity measurements were performed with a
commercial a.c.\ transport setup (quantum design measurement
system) using the conventional four-point method.

\begin{figure}
\begin{center}
\leavevmode
\psfig{figure=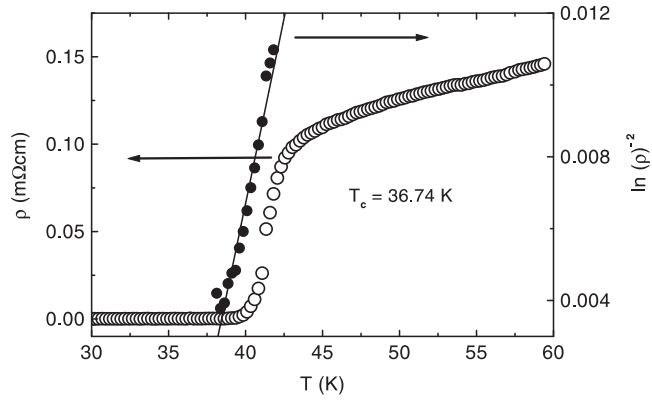,width=85mm}
\end{center}
\caption{
Temperature dependence of the resistivity $\rho$ and
$( \ln \rho ) ^{-2}$ in zero field. The solid line is a
fit to Eq.\ (\ref{eq26}) providing an estimate for the Kosterlitz--Thouless
transition temperature.
}
\label{fig2}
\end{figure}

In Fig.\ \ref{fig2} we depict the zero-field temperature
dependence of the La$_{2-x}$Sr$_{x}$CuO$_{4}$ film. Included is
the plot $(\ln \rho )^{{-2}}$ versus $T$. According to Eq.\
(\ref{eq26}) it provides the estimate $T_{c}(H=0)\approx 36.74$ K
for the Kosterlitz--Thouless transition temperature.

\begin{figure}
\begin{center}
\leavevmode
\psfig{figure=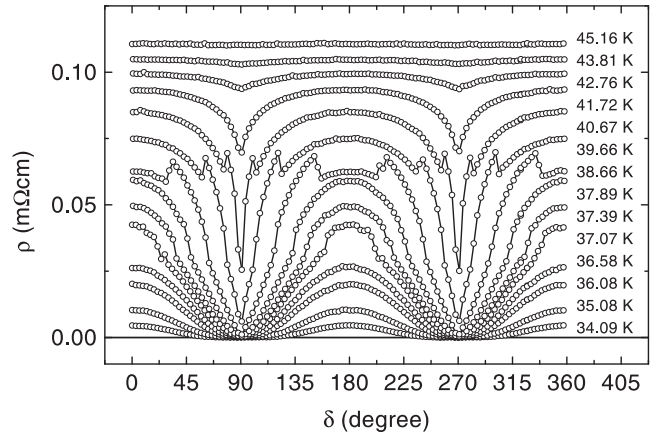,width=85mm}
\end{center}
\caption{
Angular dependence of the resistivity of a 120-\AA\/-thick
La$_{2-x}$Sr$_{x}$CuO$_{4}$ film with $T_{c}( H=0) \approx 36.74$ K
at various temperatures at $H=1$ T.
}
\label{fig3}
\end{figure}

The angular dependence of the resistivity in a magnetic field of
$H=1$ T is shown in Fig.\ \ref{fig3} for different temperatures.
The parameter $\delta$ is the angle between the magnetic field
and the $c$ axis of the film.

\begin{figure}
\begin{center}
\leavevmode
\psfig{figure=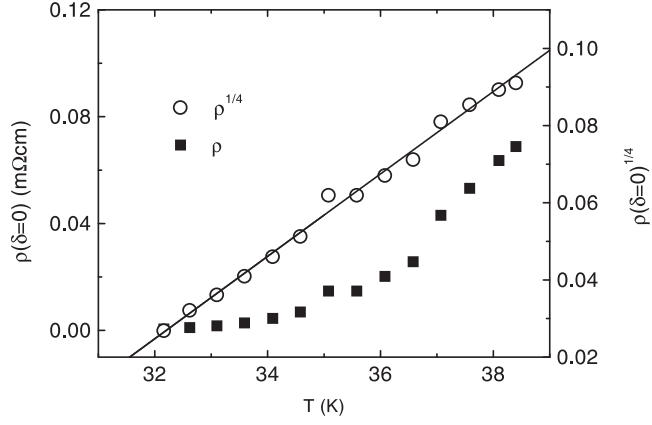,width=85mm}
\end{center}
\caption{
$\rho ( \delta =0) $ versus $T$ and $\rho (
\delta =0) ^{1/4}$ versus $T$ at $H=1$ T, yielding the
estimate $T_{g}\approx 29.7$ K.
}
\label{fig4}
\end{figure}

For fixed $\delta$, the resistivity decreases with decreasing
temperature. To provide some preliminary evidence of the
vortex--glass transition in D = 3, we show in Fig.\ \ref{fig4}
$\rho (\delta =0)$ versus $T$ and $\rho ^{1/4}$ versus $T$ [see
Eq.\ (\ref{eq7})], assuming that\cite{Mun96}
\begin{equation}
\nu _{g}(2+z_{g}-D)=\nu _{g}(z_{g}-1)\approx 4.  \label{eq35}
\end{equation}
One can see that $\rho ^{1/4}$ is linearly dependent on $T$, which yields
the estimate
\begin{equation}
T_{g}(\delta =0,H=1\;{\rm T})\approx 29.7\;{\rm K}.  \label{eq36}
\end{equation}

Below $T_{c}(H=0) =36.74$ K, where at constant field and in D = 3
the vortex--glass transition is expected to occur, the angular
dependence shown in Fig.\ \ref{fig3} appears to be consistent with
anisotropic 3D bulk behavior, whereas at higher temperatures a
crossover to 2D behavior sets in, characterized by a sharp drop of
the resistivity around 90$^{\circ}$ and 270$^{\circ}$.

\begin{figure}
\begin{center}
\leavevmode
\psfig{figure=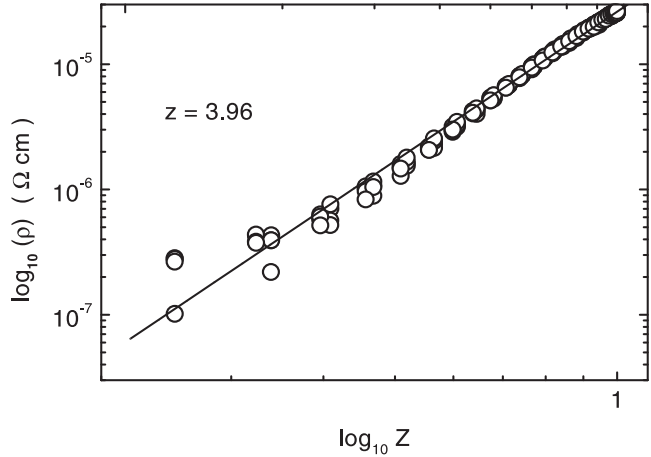,width=85mm}
\end{center}
\caption{
Log$_{10}( \rho ) $ versus $\log_{10}( {\cal Z}( H,\delta ) ) $
for $T=36.58$ K and $H=1$ T. The
solid line corresponds to a linear fit yielding  the
estimate $z=3.96\pm 0.02$, using Eq.\ (\ref{eq38}).
}
\label{fig5}
\end{figure}

To explore the 2D-$XY$ regime more quantitatively we consider
$T=36.58$ K, which is close to $T_{c}( H=0) \approx $ 36.74 K,
where the Kosterlitz--Thouless transition occurs (see Fig.\
\ref{fig2}). In this temperature regime and for sufficiently small
magnetic fields, 2D fluctuations are expected to dominate and,
according to Eq.\ (\ref{eq34}), a simple power-law behavior is
expected. Accordingly, a plot $\log_{10}\left( \rho \right) $
versus $\log ( {\cal Z}(H,\delta ))$ providing an estimate for $z$
appears to be appropriate with
\begin{equation}
(Hd^{2})/(12\Phi _{0})=0.0058  \label{eq37}
\end{equation}
for $d=120$ \AA\/ and $H=1$ T. \ The plot shown in Fig.\
\ref{fig5} points to the characteristic power law behavior valid in
D = 2 and yields the estimate
\begin{equation}
z=3.96\pm 0.02,  \label{eq38}
\end{equation}
This estimate differs from the value of $z=2$ expected from
simple diffusion, and is closer to $z= 5.6$, the value that
emerges from a recent re-analysis of the experimental data for 2D
superconductors, Josephson-junction arrays, and
superfluids.\cite{Pierson99}

\begin{figure}
\begin{center}
\leavevmode
\psfig{figure=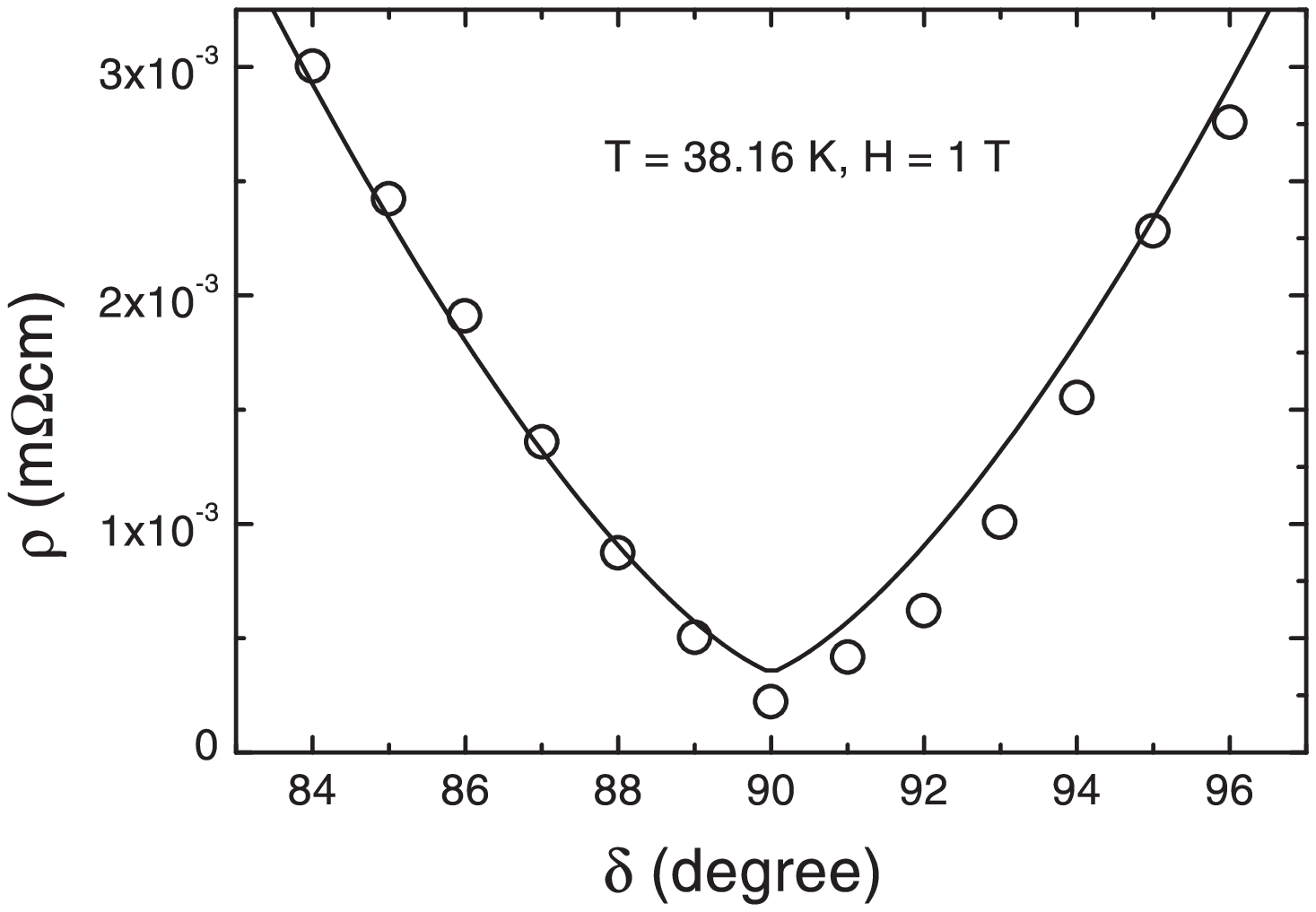,width=85mm}
\end{center}
\caption{
$\rho$ versus $\delta$ for $T=38.16$ K and $H=1$ T. The
solid line is a guide to the eye to emphasize the characteristic
2D behavior (V-shape).
}
\label{fig6}
\end{figure}

\begin{figure}
\begin{center}
\leavevmode
\psfig{figure=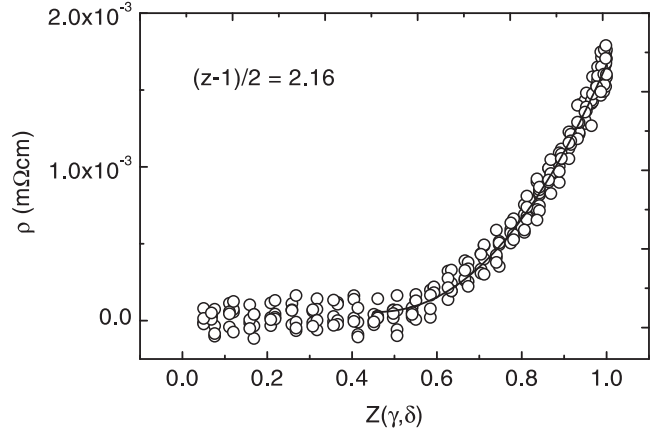,width=85mm}
\end{center}
\caption{
$\rho$ versus ${\cal Z}( \gamma ,\delta ) $
for $T=33.1$ K and $H=1$ T. The solid line is a fit to Eq.\ (\ref{eq42}),
yielding the parameters listed in Table \ref{tableI}.
}
\label{fig7}
\end{figure}

\begin{figure}
\begin{center}
\leavevmode
\psfig{figure=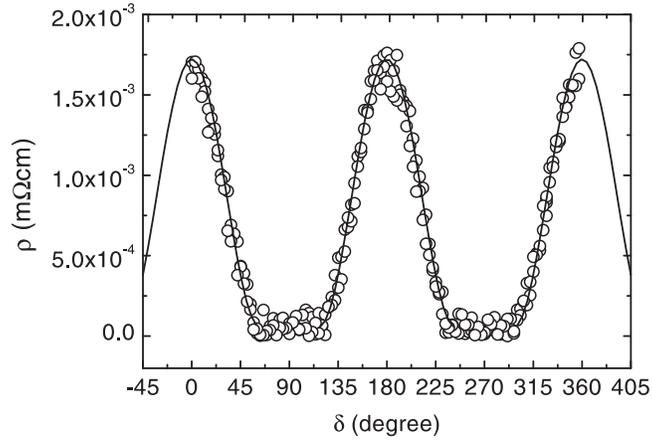,width=85mm}
\end{center}
\caption{
$\rho$ versus $\delta$ for $T=33.1$ K and $H=1$ T. The
solid line corresponds to Eq.\ (\ref{eq42}), with the parameters listed in
Table \ref{tableI}.
}
\label{fig8}
\end{figure}

Another manifestation of dominant 2D fluctuations, requiring $\xi
_{c}\equiv \xi _{\bot }\geq d$, is shown in Fig.\ \ref{fig6} in
terms of $\rho$ versus $\delta$ around $\delta =90^{\circ }$ for
$T=38.16$ K and $H=1$ T. Indeed, the characteristic V-shape
clearly indicates the 2D scaling behavior in this temperature
regime. From Fig.\ \ref{fig3} it is seen, however, that the 2D
behavior disappears as the temperature is lowered. In this
context it should be noted that there are three important
length scales in the problem:\ (i) the average distance between
the vortices
\begin{equation}
l=\left( \frac{\Phi _{0}}{\pi H}\right) ^{1/2}\approx 256\,{\rm \AA
\/\;at\;1\;T};  \label{eq39}
\end{equation}
(ii) the correlation length parallel to the layers, $\xi _{\Vert}$.
The critical regime for the 3D vortex glass-to-normal
transition requires, at the very least, that
\begin{equation}
\xi _{\Vert }\geq l,  \label{eq40}
\end{equation}
which is thus more accessible in large fields; and (iii) the
correlation length perpendicular to the layers, $\xi _{\bot }$. In
experiments on films of thickness $d$, 3D critical behavior
requires [Eq.\ (\ref{eq22})] that
\begin{equation}
\xi _{\bot }=\frac{\xi _{\Vert }}{\gamma }<d.  \label{eq41}
\end{equation}
The absence of the characteristic 2D V-shape in the angular
dependence of the resistivity around $\delta =90^{\circ}$ and
270$^{\circ}$ and below $T=36$ K (see Fig.\ \ref{fig3}) suggests
that in the regime considered here, condition (\ref{eq40}) is
satisfied. Indeed, from Fig.\ \ref{fig7} it is seen that the
resistivity appears to vanish below a certain value of the
scaling variable. This suggests that in this temperature regime
the 3D scaling form (\ref{eq20}) for the normal-to-vortex glass
transition applies, so that
\begin{equation}
\rho =a+b({\cal Z}(\gamma ,\delta )-\widetilde{{\cal
Z}}_{g})^{(z_{g}-1)/2}, \ \widetilde{{\cal Z}}_{g}=\frac{{\cal
Z}_{g}}{{\cal Z}(H,T)}  \label{eq42}
\end{equation}
where ${\cal Z}(\gamma ,\delta )$ is given by Eq.\ (\ref{eq14}) and
$a$ accounts for the residual resistivity. Here we used $\gamma
\approx 20$, the value derived from magnetic torque measurements
on bulk samples close to optimum doping.\cite{Hofer99} The fit
parameters are listed in Table \ref{tableI}. At $\widetilde{{\cal
Z}}_{g}\approx 0.45$ the resistivity approaches zero and vanishes
below this threshold, signaling the vortex-glass phase. The
resulting behavior of the angular-dependent resistivity is
depicted in Fig.\ \ref{fig8}. Here the vortex-glass phase appears in the
interval $65.95\leq \delta \leq 114.05^{\circ }$ (modulo $\pi $).
In Table \ref{tableI} we also included the estimates derived from
the data taken at $T=33.1$, 33.6 and 34.1 K. The consistency of
the estimates for $(z_{g}-1)/2$ and ${\cal Z}_{g}$ suggests that
in the temperature regime considered here the 3D normal
phase-to-vortex glass transition was attained at constant
temperature and magnetic field by varying the orientation of the
magnetic field.

\begin{figure}
\begin{center}
\leavevmode
\psfig{figure=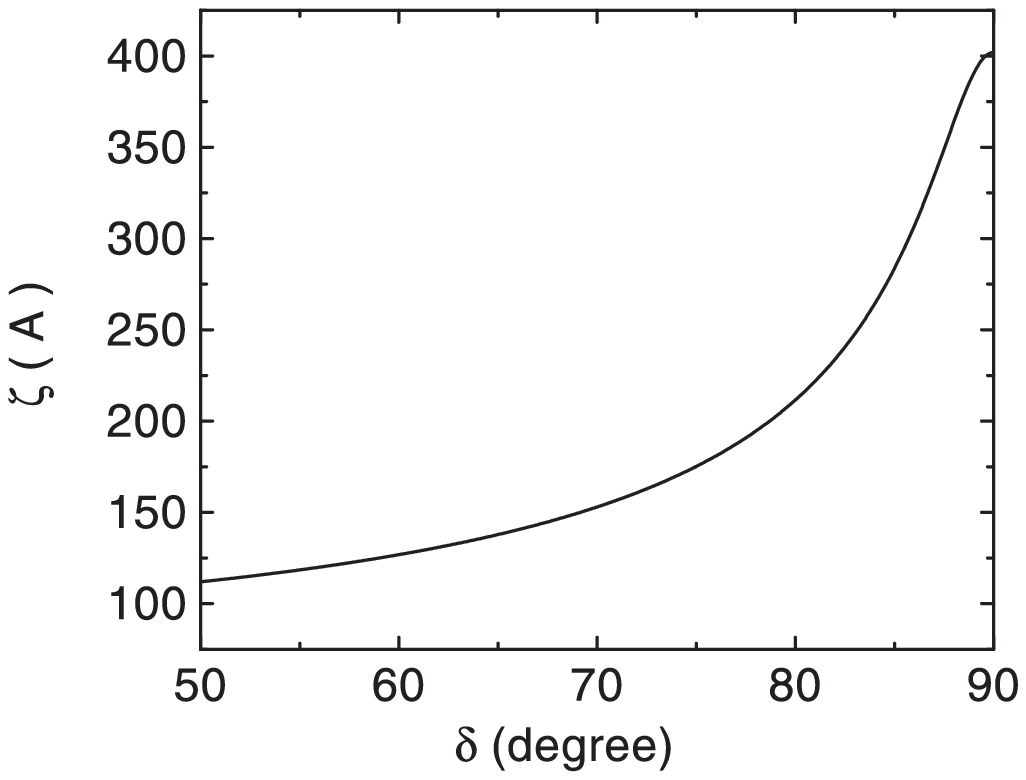,width=85mm}
\end{center}
\caption{
Contour plot of ${\cal Z}( \xi _{\Vert },\delta
) =0.039$ [Eq.\ (\ref{eq43})] for $H=1$ T.
}
\label{fig9}
\end{figure}

Substantiation of this conclusion amounts to show that ${\cal
Z}_{g}$ can be approached without violating the condition given by
Eqs.\ (\ref{eq40}) and (\ref{eq41}). A rough estimate for
${\cal Z}_{g}$ is obtained using Eq.\ (\ref{eq17}) with $\xi
_{\Vert ,0}\approx 30$ \AA\/, $T_{c}(H=0)\approx 36.74 $ K and
$T_{g}(H=1\,{\rm T})\approx 29.7$ K [Eq.\ (\ref{eq36})], yielding
${\cal Z}_{g}\approx 0.039$. On the other hand, from ${\cal Z}
(H,T)=(\xi _{\Vert }^{2}H)/\Phi _{0}$ we find ${\cal Z}(H=1\,{\rm
T} ,\,T=33.1\,{\rm K})\approx 0.094$, using $\xi _{\Vert }=\xi
_{\Vert ,0}(1-33.1/36.74)^{-2/3}\approx 140$ \AA\/ for $\xi
_{\Vert ,0}\approx 30$ \AA \/. Hence $\ \widetilde{{\cal
Z}}_{g}(T=33.1\,{\rm K})={\cal Z}_{g}/{\cal Z} (H,T)\approx 0.41$,
in remarkable agreement with the value listed in Table
\ref{tableI}. Moreover, considering
\begin{equation}
{\cal Z}(\xi _{\Vert },\delta )={\cal Z}(H=1T,T){\cal Z}(\gamma
,\delta )= {\cal Z}_{g}=0.039  \label{eq43}
\end{equation}
with ${\cal Z}(H,T)$ and ${\cal Z}(\gamma ,\delta )$ given by Eq.\
(\ref{eq14}), it is readily seen from Fig.\ \ref{fig9} that in the
interval $80^{\circ }\leq \delta \leq 110^{\circ }$, ${\cal
Z}_{g}$ can be approached without violating condition
(\ref{eq40}). Moreover, noting that $\xi _{\bot }=\xi _{\Vert
}/\gamma \approx 7<d=120$ \AA\/ ($\gamma =20$), constraint
(\ref{eq41}) turns out to be much less stringent. On this basis
we conclude that the estimates listed in Table \ref{tableI} and
the data shown in Figs.\ \ref{fig7} and \ref{fig8} reveal the
critical behavior of the 3D normal-to-vortex glass transition.
The dynamic critical exponent is close to
\begin{equation}
z_{g}\approx 5.7,  \label{eq44}
\end{equation}
in reasonable agreement with previous estimates:\ $z_g=4.8$ for
YBa$_{2}$Cu$_{3}$O$_{7-\delta }$ films,\cite{Fisher91a}
$z_{g}=4.4\pm 0.1$ for YBa$_{2}$Cu$_{3}$O$_{7-\delta }$
films,\cite{Kotzler94} $z_{g}=4.3$ for YBa$_{2}$Cu$
_{3}$O$_{7-\delta }$ single crystals,\cite{Koch89} $z_{g}=5.46\pm
0.32$ for YNi$_{2}$B$_{2}$C single crystals,\cite{Woltgens95} and
$z_{g}\approx 4.85$ for optimally doped
YBa$_{2}$Cu$_{3}$O$_{7-\delta }$ films.\cite{Rydh99} It differs
markedly, however, from the value at $T=T_{c}(H=0)$ and $H=1$ T,
$z\approx 3.96$ [Eq.\ (\ref{eq38})], where 2D-$XY$ fluctuations
dominate.

To summarize, we have shown that the angular dependence of the
resistivity measured at various temperatures in an applied
magnetic field of fixed magnitude allows us to extract various
critical properties of interest, including the 2D-to-3D crossover
and the 3D vortex glass-to-fluid transition as the temperature is
decreased. The data turned out to be remarkably consistent with the
scaling theory for the d.c.\ resistivity of an anisotropic
superconductor in a magnetic field applied at an angle $\delta $
to the $c$ axis. Moreover, we have shown that the presence or
absence of the V-shaped angular dependence of the resistivity
around $\delta =90^{\circ }$ (modulo $\pi$) provides an
unambiguous tool to determine the effective dimensionality of the
system as the temperature is decreased in a constant magnetic field.

\acknowledgments
We thank J.\ M.\ Triscone and {\O
}.\ Fisher for useful discussions, and the Swiss National Science
Foundation for partial support.

\clearpage
\begin{table}[tbp]
\caption{Fit parameters of Eq.\ (\ref{eq32}); note that
$b^{2/(z-1)}\propto {\cal Z}(H,T)$ and $\widetilde{{\cal
Z}}_{g}b^{2/(z-1)}\propto {\cal Z}_{g}$ . }
\label{tableI}
\begin{tabular}{ccccccc}
$T$ (K) & $\widetilde{{\cal Z}}_{g}$ & $(z-1)/2$ & $b$ &
$b^{2/(z-1)}$ & $ \widetilde{{\cal Z}}_{g}b^{2/(z-1)}$ & $a$
\\ \hline
$33.1$ & $0.41$ & $2.33\pm 0.07$ & $5.55 \times 10^{-6}$ &
$5.55 \times 10^{-3}$ & $2.28 \times 10^{-3}$ & $4.06 \times 10^{-8}$
\\
$33.6$ & $0.38$ & $2.36\pm 0.04$ & $8.41 \times
10^{-6}$ & $7.07 \times 10^{-3}$ & $2.69 \times 10^{-3}$ & $1.17 \times 10^{-7}$
\\
$34.1$ & $0.33$ & $2.34\pm 0.02$ & $1 \times
10^{-5}$ & $7.30 \times 10^{-3}$ & $2.41 \times 10^{-3}$ & $1.47 \times 10^{-7}$
\end{tabular}
\end{table}

\clearpage

\bigskip

\end{document}